

\magnification = 1200
\overfullrule=0pt

\font\titlerm = cmr10 scaled\magstep 4
\font\titlerms = cmr7 scaled\magstep 4
\font\titlermss = cmr5 scaled\magstep 4
\font\titlei = cmmi10 scaled\magstep 4
\font\titleis = cmmi7 scaled\magstep 4
\font\titleiss = cmmi5 scaled\magstep 4
\font\titlesy = cmsy10 scaled\magstep 4
\font\titlesys = cmsy7 scaled\magstep 4
\font\titlesyss = cmsy5 scaled\magstep 4
\font\titleit = cmti10 scaled\magstep 4

\def\titlefont{\def\rm{\fam0\titlerm}
\def\it{\fam\itfam\titleit}
\textfont0 = \titlerm
\scriptfont0 = \titlerms
\scriptscriptfont0 = \titlermss
\textfont1 = \titlei
\scriptfont1 = \titleis
\scriptscriptfont1 = \titleiss
\textfont2 = \titlesy
\scriptfont2 = \titlesys
\scriptscriptfont2 = \titlesyss
\textfont\itfam = \titleit
\rm}

\def\sectionfont{\def\rm{\fam0\tenrm}
\def\it{\fam\itfam\tenit}
\def\bf{\fam\bffam\tenbf}
\textfont0 = \tenrm
\scriptfont0 = \sevenrm
\scriptscriptfont0 = \fiverm
\textfont1 = \teni
\scriptfont1 = \seveni  \scriptscriptfont1=\fivei
\textfont2 = \tensy
\scriptfont2 = \sevensy
\scriptscriptfont2 = \fivesy
\textfont\itfam = \tenit
\textfont\bffam = \tenbf
\rm}

\font\teenyfont = cmr5

\global\baselineskip = 1.2\baselineskip
\global\parskip = 4pt plus 0.3pt
\global\nulldelimiterspace = 0pt



\def\endignore{}
\def\ignore #1\endignore{}

\newcount\dflag
\dflag = 0


\def\monthname{\ifcase\month
\or Jan\ \or Feb\ \or Mar\ \or Apr\ \or May\ \or June\ %
\or July\ \or Aug\ \or Sept\ \or Oct\ \or Nov\ \or Dec\ 
\fi}




\def\endid{}
\def\id#1\endid{11 May 1994
\hfill #1}

\def\endtitle{}
\def\title#1\endtitle{\vskip.15in\titlefont
\global\baselineskip = 2\baselineskip
#1\vskip.3in
\baselineskip = 0.5\baselineskip\sectionfont}

\def\lblfoot{This work was supported by the Director, Office of Energy
Research, Office of High Energy and Nuclear Physics, Division of High
Energy Physics of the U.S. Department of Energy under Contract
DE-AC03-76SF00098.}

\def\endauthors{}
\def\authors#1\endauthors{
#1\if\dflag = 0
\footnote{}{\noindent\lblfoot}\fi}

\def\endabstract{}
\def\abstract#1\endabstract{\vskip .2in%
\centerline{\sectionfont\bf Abstract}%
\vskip .1in%
\noindent#1%
\ifnum\dflag = 0
\footline = {\hfil}\pageno = 0
\vfill\eject
\pageno = 1\footline{\centerline{\sectionfont\folio}}
\fi\ifnum\dflag = 2
\footline = {\hfil}\pageno = 0
\fi}

\newcount\nsection
\newcount\nsubsection

\def\section#1{\global\advance\nsection by 1
\global\nsubsection = 0
\bigskip\noindent
\centerline{\sectionfont\bf\number\nsection.\ #1}
\medskip\sectionfont\par\nobreak}

\def\subsection#1{\global\advance\nsubsection by 1
\bigskip\noindent
\centerline{\sectionfont \it \number\nsection.\number\nsubsection.\ #1}
\medskip\rm\par\nobreak}

\def\appendix#1#2{\bigskip\noindent%
\centerline{\sectionfont \bf Appendix #1.\ #2}
\medskip\rm\par\nobreak}


\newcount\nref
\global\nref = 1

\def\ref#1#2{\xdef #1{[\number\nref]}
#1
\ifnum\nref = 1\global\xdef\therefs{\noindent[\number\nref] #2\ }
\else
\global\xdef\oldrefs{\therefs}
\global\xdef\therefs{\oldrefs\vskip.1in\noindent[\number\nref] #2\ }%
\fi%
\global\advance\nref by 1
}

\def\listrefs{\vfill\eject\section{References}\therefs}


\newcount\cflag
\newcount\nequation
\global\nequation = 1
\def\eqlabel{(1)}

\def\nexteqno{\ifnum\cflag = 0
\global\advance\nequation by 1
\fi
\global\cflag = 0
\xdef\eqlabel{(\number\nequation)}}

\def\lasteqno{\global\advance\nequation by -1
\xdef\eqlabel{(\number\nequation)}}

\def\label#1{\xdef #1{(\number\nequation)}
\ifnum\dflag = 1
{\escapechar = -1
\xdef\draftname{\teenyfont\string#1}}
\fi}

\def\clabel#1#2{\xdef\eqlabel{(\number\nequation #2)}
\global\cflag = 1
\xdef #1{\eqlabel}
\ifnum\dflag = 1
{\escapechar = -1
\xdef\draftname{\string#1}}
\fi}

\def\cclabel#1#2{\xdef\eqlabel{#2)}
\global\cflag = 1
\xdef #1{\eqlabel}
\ifnum\dflag = 1
{\escapechar = -1
\xdef\draftname{\string#1}}
\fi}


\def\eeq{}

\def\eqnn #1\eeq{$$ #1 $$}

\def\eq #1\eeq{\xdef\draftname{\ }
$$ #1
\eqno{\eqlabel \rlap{\ \draftname}} $$
\nexteqno}



\def\eol{& \eqlabel \rlap{\ \draftname} \crcr
\nexteqno
\xdef\draftname{\ }}

\def\eeol{& \eqlabel \rlap{\ \draftname}
\nexteqno
\xdef\draftname{\ }}



\def\eqa #1\eeq{\xdef\draftname{\ }
$$ \eqalignno{ #1 } $$
\global\cflag = 0}


\newcount\nfig
\global\nfig = 1

\def\fg#1\efig{\vskip .5in\noindent Fig.\ \number\nfig:\ #1%
\global\advance\nfig by 1}



\def\eg{{\it e.g.\/}}

\def\etc{{\it etc}}

\def\myinstitution{
    \centerline{\it Theoretical Physics Group}
    \centerline{\it Lawrence Berkeley Laboratory}
    \centerline{\it 1 Cyclotron Road}
    \centerline{\it Berkeley, California 94720}
}


\def\jref#1#2#3#4{{\it #1} {\bf #2}, #3 (#4)}

\def\NPB#1#2#3{\jref{Nucl.\ Phys.}{B#1}{#2}{#3}}

\def\PLB#1#2#3{\jref{Phys.\ Lett.}{#1B}{#2}{#3}}

\def\PRD#1#2#3{\jref{Phys.\ Rev.}{D#1}{#2}{#3}}

\def\PRL#1#2#3{\jref{Phys.\ Rev.\ Lett.}{#1}{#2}{#3}}


\def\to{\mathop{\rightarrow}}


\def\myint{\int\mkern-5mu}
\def\frac#1#2{{{#1} \over {#2}}\,}  


\def\Dsl{\hbox{/\kern-.6000em\rm D}} 



\def\twi{\widetilde}

\def\scr#1{{\cal #1}}
\def\op#1{{\widehat #1}}
\def\mybar#1{\kern 0.8pt\overline{\kern -0.8pt#1\kern -0.8pt}\kern 0.8pt}
\def\sla#1{\raise.15ex\hbox{$/$}\kern-.57em #1}
\def\Sla#1{\kern.15em\raise.15ex\hbox{$/$}\kern-.72em #1}

\def\roughly#1{\mathrel{\raise.3ex\hbox{$#1$\kern-.75em%
    \lower1ex\hbox{$\sim$}}}}


\def\tr{\mathop{\rm tr}}





\hyphenation{ba-ry-on ba-ry-ons ano-ma-ly ano-ma-lies}

\def\al{\alpha}
\def\del{\delta}
\def\Del{\Delta}
\def\gam{\gamma}
\def\Gam{\Gamma}
\def\ep{\epsilon}
\def\lam{\lambda}
\def\Lam{\Lambda}

\def\Om{\Omega}
\def\sig{\sigma}
\def\Sig{\Sigma}

\def\ChPT{\raise.45ex\hbox{$\chi$}PT}

\def\lhs{left-hand side}

\def\hc{{\rm h.c.}}


\def\MeV{{\rm \ MeV}}
\def\GeV{{\rm \ GeV}}

\hyphenation{ba-ry-on ba-ry-ons}

\def\chisim{$SU(N_F)_L \times SU(N_F)_R$}

\def\rbra#1{(#1|}
\def\rket#1{|#1)}

\def\up{\uparrow}
\def\down{\downarrow}

\def\op#1{\mathord{\{#1\}}}

\def\mn{\mu^{\vphantom j}_N}


\id
LBL-35598, CfPA-94-TH-25
\endid
\rightline{hep-ph/9405272}

\vskip -.15in
\title
\centerline{Baryon Magnetic Moments in a}
\centerline{Simultaneous Expansion in $1/N$ and $m_s$}
\endtitle

\authors
\centerline{Markus A. Luty\ \ {\it and}
\ \ John March--Russell${}^*$\footnote{}
{\hskip-.27in ${}^*$ DOE distinguished postdoctoral research fellow.}}
\vskip .1in
\myinstitution
\vskip .15in
\centerline{Martin White${}^\dagger$\footnote{}
{\hskip-.27in ${}^\dagger$ SSC Fellow.}}
\vskip .1in
\centerline{{\it Center for Particle Astrophysics}}
\centerline{{\it 301 Le Conte Hall}}
\centerline{{\it University of California}}
\centerline{{\it Berkeley, CA 94720}}
\footnote{}{\hskip-.27in\lblfoot}
\endauthors

\abstract
We consider the baryon octet and decuplet magnetic moments in a simultaneous
expansion in $m_s$ and $1/N$ taking $N_F / N \sim 1$, where $N$ is the
number of QCD colors and $N_F$ is the number of light quark flavors.
At leading order in this expansion, the magnetic moments obey the
non-relativistic quark-model relations.
We compute corrections to these relations using an effective lagrangian
formalism which respects chiral symmetry to all orders in the $1/N$ expansion.
Including corrections up to order $m_s^{1/2}$, we find 8 relations among the
9 measured octet and decuplet magnetic moments;
including corrections up to order $1/N$ and $m_s$, we find 4 remaining
relations.
The relations work well, and suggest that the expansion is under control.
We give predictions for the unmeasured decuplet magnetic moments.
\endabstract


\section{Introduction}

In this paper, we consider the baryon magnetic moments in a simultaneous
expansion in $m_s$ and $1/N$, where $N$ is the number of QCD colors.
There has been a recent revival of interest in the $1/N$ expansion for
baryons, started by the results of
ref.\ \ref\UCSD{R. Dashen and A. V. Manohar, \PLB{315}{425}{1993};
\PLB{315}{438}{1993}; E. Jenkins, \PLB{315}{441}{1993}.}.
For example, it was shown that many interesting large-$N$ relations have
corrections starting at $\sim 1/N^2$.
These results have been extended using a number of different methods
\ref\bigUCSD{R. Dashen, E. Jenkins, and A. V. Manohar, UCSD/PTH 93-21,
hep-ph/9310379.}
\ref\Harvard{C. Carone, H. Georgi, and S. Osofsky, HUTP-93/A032,
hep-ph/9310365.}
\ref\us{M. A. Luty and J. March--Russell, LBL-34778, hep-ph/9310369,
submitted to {\it Nucl.\ Phys.} {\bf B}.}
\ref\largeNF{M. A. Luty, LBL-35539, NSF-ITP-94-42, hep-ph/9405271}.
We will use the formalism of ref.\ \us.
This formalism is based on an exact relativistic treatment of QCD, yet
makes direct contact with the static quark model.
On a more practical level, it allows us to write an explicit effective
lagrangian in which chiral symmetry is kept manifest to all orders in $1/N$.

When the number of light flavors $N_F > 2$, the $SU(N_F)$ flavor
representations of baryons grow with $N$, and so there are ambiguities in
how to extrapolate physical baryon states to $N > 3$.
We use the approach of ref.\ \largeNF, where it is shown that the $1/N$
expansion can be formulated to include all of the states in the $SU(N_F)$
flavor representations for arbitrary $N$ and $N_F$.
The physical results for $N = 3$ can be obtained without having to identify
the physical baryon states with particular states for $N > 3$, and $SU(N_F)$
flavor symmetry is also kept manifest in this approach.
This expansion is well-defined even if $N_F / N \sim 1$ \largeNF;
we will work in this limit, since $N_F = N = 3$ in the real world.

In the large-$N$ limit, the magnetic moments obey the non-relativistic
quark model relations
\ref\Skyrme{A. V. Manohar, \NPB{248}{19}{1984}.}.
We find that the leading corrections to these results are suppressed
relative to the leading terms by order $1/N$, $m_s^{1/2}$, and $m_s$.
We will assume that $O(m_s)$ and $O(1/N)$ corrections are both
$O(\ep) \sim 30\%$, and carry out the expansion consistently to $O(\ep)$.
Including the $O(m_s^{1/2}) = O(\ep^{1/2})$ corrections, we find 8 relations
among the 9 measured magnetic moments for the octet and decuplet baryons;
including corrections up to $O(\ep)$, we find 4 surviving relations.
These relations agree well with data, providing evidence that the combined
$1/N$ and chiral expansions work well for baryons.
We then predict the unmeasured decuplet magnetic moments including all
corrections up to $O(\ep)$; these may be measured at CEBAF.

\section{Formalism}

In this section, we briefly review the formalism we will use to obtain our
results.
In ref.\ \us, it was shown how to write an effective lagrangian describing
the low-energy interactions of baryons with the pseudo-Nambu--Goldstone
bosons (PNGB's) for large $N$.
The PNGB's are described in the standard way:
the field
\eq
\xi(x) = e^{i\Pi(x) / f},
\eeq
is taken to transform under \chisim\ as
\eq
\label\Udef
\xi \mapsto L \xi U^\dagger = U \xi R^\dagger,
\eeq
where this equation implicitly defines $U$ as a function of $L$, $R$,
and $\xi$.
For $N_F = 3$, the meson fields are
\eq
\Pi = \frac 1{\sqrt 2}
\pmatrix{\frac 1{\sqrt 2}\pi^0 + \frac 1{\sqrt 6}\eta &
\pi^+ & K^+ \cr
\pi^- & -\frac 1{\sqrt 2} \pi^0 + \frac 1{\sqrt 6}\eta &
K^0 \cr
K^- & {\mybar K}^0 & -\frac 2{\sqrt 6} \eta \cr}.
\eeq
Note that the $\eta'$ is not light if $N_F / N \sim 1$, and is therefore
not included.
The effective lagrangian is most conveniently written in terms of the
hermitian fields
\eq
V_\mu \equiv \frac i2\left(\xi \partial_\mu \xi^\dagger
+ \xi^\dagger \partial_\mu \xi\right), \qquad
A_\mu \equiv \frac i2\left(\xi \partial_\mu \xi^\dagger
-\xi^\dagger \partial_\mu \xi\right),
\eeq
which transform under \chisim\ as
\eq
V_\mu \mapsto U V_\mu U^\dagger + iU\partial_\mu U^\dagger, \qquad
A_\mu \mapsto U A_\mu U^\dagger.
\eeq

We can incorporate $SU(N_F)$ breaking due to $m_s \neq 0$ by including the
quark mass spurion (for arbitrary $N_F$)
\eq
\label\qmass
m_q \equiv m_s S, \quad
S \equiv \pmatrix{0 &&&\cr & \ddots &&\cr && 0 &\cr &&& 1 \cr},
\eeq
transforming under \chisim\ as $m_q \mapsto L m_q R^\dagger$.
We find it convenient to define the even-parity field
\eq
m \equiv \frac 12 (\xi^\dagger m_q \xi + \hc)\mapsto U m U^\dagger.
\eeq
In this notation, the leading terms giving rise to meson interactions are
\eq
\label\mesonL
\scr L = f^2 \tr(A^\mu A_\mu) + \beta f^3 \tr(m) + \cdots,
\eeq
where $\beta$ is a coupling and $f = f_\pi \simeq 93 \MeV$;
$f \sim \sqrt{N}$ in the large-$N$ limit
\ref\Witten{See \eg\ E. Witten, \NPB{160}{57}{1979}.}.

We now discuss the baryon fields.
Because the baryon mass is of order $N \Lam_{\rm QCD}$, we can describe
the baryons using a heavy-particle effective field theory
\ref\heavyB{E. Jenkins and A. V. Manohar, \PLB{255}{558}{1991}.}.
We write the baryon momentum as $P = M_0 v + k$, where $M_0 \sim N$ is a
baryon mass and $v$ is a 4-velocity ($v^2 = 1$) which defines the baryon
rest frame.
We then write an effective field theory in terms of baryon fields whose
momentum modes are the residual momenta $k$.
This effective field theory gives an expansion in $1/M_0$ around the
static limit.

For $N$ large, the baryon $SU(N_F)$ representations are large, and it is
convenient to use a compact notation to keep track of baryon flavor quantum
numbers.
We use a Fock-space notation in which the baryons fields are written
\eq
\rket{\scr B(x)} \equiv \scr B^{a_1 \al_1 \cdots a_N \al_N}(x)\,
\al^\dagger_{a_1 \al_1} \cdots \al^\dagger_{a_N \al_N} \rket 0.
\eeq
The $\al^\dagger$'s are {\it bosonic} creation operators which create a
``quark'' with definite flavor and spin, and
$\rket 0$ is the Fock ``vacuum'' state;
$a_1, \ldots, a_N$ are $SU(N_F)$ flavor indices and
$\al_1, \ldots, \al_N = \up, \down$ are spin indices in the rest frame
defined by $v$.
\ignore
(In a fully relativistic notation, $\al_1, \ldots, \al_N$ would be Dirac
indices running from 1 to 4, and the fields $\scr B$ would obey the
constraints
$(\sla v)^{\al_j}_\beta \scr B^{a_1 \al_1 \cdots a_j \beta \cdots a_N \al_N} =
\scr B^{a_1 \al_1 \cdots a_N \al_N}$, $j = 1,\ldots, N$.)
\endignore

Under \chisim, the baryon fields transform as
\eq
\scr B^{a_1 \al_1 \cdots a_N \al_N} \mapsto
{U^{a_1}}_{b_1} \cdots {U^{a_N}}_{b_N} \scr B^{b_1 \al_1 \cdots b_N \al_N},
\eeq
where $U$ is defined in eq.\ \Udef.
$\scr B^{\cdots}$ transforms under a highly reducible representation of
$SU(N_F)$, but we will have to carry out calculations explicitly only for
$N = N_F = 3$.

Meson--baryon interactions are written in terms of operators constructed
from the creation and annihilation operators.
For example, the leading terms involving baryons can be written
\eq
\label\lzero
\scr L = \rbra{\scr B} iv^\mu \nabla_\mu \rket{\scr B}
+ g\, \rbra{\scr B} \op{A^\mu \sig_\mu} \rket{\scr B} + \cdots,
\eeq
where $\sig_\mu \equiv (\sla v \gam_\mu - v_\mu) \gam_5$ is the
spin matrix and we define
\eq
\op{A^\mu \sig_\mu} \equiv \al^\dagger_{a\al}
{(A^\mu)^a}_b {(\sig_\mu)^\al}_\beta \al^{b\beta},
\eeq
\etc.
The chiral covariant derivative acting on the baryon fields is defined by
\eq
\nabla_\mu \rket{\scr B} \equiv (\partial_\mu - i\op{V_\mu}) \rket{\scr B}.
\eeq
The coupling $g$ can be determined from matrix elements of the $\Del S = 1$
axial current measured in semileptonic hyperon decays.
We obtain $g = 0.83 \pm 0.08$, where the error is obtained by assigning a
$30\%$ uncertainty to the higher-order corrections.
This value should be used with caution, since the $SU(N_F)$-breaking
corrections are known to be large \heavyB
\ref\bigcorr{M. A. Luty and M. White, \PLB{319}{261}{1993}}.

In this notation, the leading $N$ dependence of an arbitrary term in the
effective lagrangian is given by associating a factor of $1/N^{r - 1}$
with every $r$-body operator (that is, an operator constructed from $r$
creation and $r$ annihilation operators), and a factor of $1/N$ for every
explicit flavor trace \us.
The reason for these rules is that an $r$-body operator can arise only
from quark-level diagrams involving at least $r - 1$ gluon exchanges, and
flavor traces arise from quark loops.
Each gluon exchange or quark loop gives rise to a suppression of $1/N$,
yielding the rules given above.
For more details, see ref.\ \us.
According to these rules, the coupling $g$ in eq.\ \lzero\ is order 1 in the
large-$N$ limit.

\section{Magnetic Moments}

We now apply the formalism discussed in the previous section to the
magnetic moments.

\subsection{Leading Order}

At leading order in the large-$N$ and chiral limits, the baryon magnetic
moments are described by a single term in the effective lagrangian:
\eq
\label\leading
\del\scr L = \frac{a_0 e}\Lam\, v_\mu \twi F^{\mu\nu}
\rbra{\scr B} \op{\scr Q \sig_\nu} \rket{\scr B},
\eeq
where $\twi F^{\mu\nu} \equiv \frac 12 \ep^{\mu\nu\lam\rho}
F_{\lam\rho}$, $\ep_{0123} = +1$, and
\eq
\scr Q \equiv
\frac 12 (\xi^\dagger \scr Q_L \xi^\dagger +  \xi \scr Q_R \xi)
\mapsto U \scr Q U^\dagger.
\eeq
Here (at $N_F = 3$)
\eq
\scr Q_L = \scr Q_R \equiv
\pmatrix{\frac 23 &&\cr & -\frac 13 &\cr && -\frac 13 \cr}
\eeq
are the left- and right-handed quark charge spurions.
The parameter $\Lam \sim 1 \GeV$ is the chiral expansion scale;
na\"\i ve dimensional analysis leads us to expect that $a_0 \sim 1$.
The magnetic moments arising from eq.\ \leading\ are given by
\eq
\mu^{\vphantom j}_{B' B}\, \sig^j_{B' B} =
-\frac{a_0 e}\Lam\, \rbra{B'} \op{\scr Q \sig^j} \rket{B},
\eeq
where $\sig^j_{B' B}$ is the matrix element of the spin matrix
between the states $\rket{B}$ and $\rket{B'}$ normalized so
that its maximal value is $+1$.
(We use non-script capital letters to refer to specific baryon states.)
The operator $\op{\scr Q \sig^3}$ has matrix elements $O(N)$, so that
the leading contributions to the magnetic moments are
$O(N)$.\footnote{$^*$}
{We could normalize the electric charge of the quarks to be of order $1/N$
so that the baryons have electric charge of order 1 in the large-$N$ limit.
If we did this, the magnetic moments would be $O(1)$ in the large-$N$
limit.
Such a rescaling would modify the formulas which follow in a trivial way,
and would not affect our results for $N = 3$.}

At this order, there are 9 measured octet and decuplet baryon magnetic
moments determined by a single unknown constant $a_0 / \Lam$.
There are therefore 8 relations among the magnetic moments.
Of these, 6 are the Coleman--Glashow relations
\ref\CG{S. Coleman and S. L. Glashow, \PRL{6}{423}{1961}.}
\eq
\label\lorels
\eqalign{
\mu_p - \mu_{\Sig^+} &= 0, \quad (15\%) \cr
\mu_n + \mu_p + \mu_{\Sig^-} &= 0, \quad (10\%) \cr
\mu_n - 2\mu_\Lam &= 0, \quad (40\%) \cr
\mu_{\Sig^-} - \mu_{\Xi^-} &= 0, \quad (55\%) \cr
\mu_n - \mu_{\Xi^0} &= 0, \quad (40\%) \cr
\sqrt 3 \mu_n + 2 \mu_{\Sig_0 \Lam} &= 0, \quad (5\%) \cr}
\eeq
which hold in the limit of exact $SU(3)$ independent of the $1/N$
expansion;
the remaining relations can be taken to be the quark-model relations
\eqa
\label\QM
3 \mu_n + 2 \mu_p &= 0, \quad (3\%) \eol
\label\QMtoo
\mu_{\Om^-} + \mu_p &= 0, \quad (35\%) \eeol
\eeq
which are consequences of the large-$N$ limit.
The numerical accuracy indicated is defined by dividing the numerical
value by the average of the positive and negative terms on the \lhs.
If we perform a best fit, the average deviation is $0.3\ \mn$.

\subsection{$1/N$ Corrections}

The $O(1/N)$ corrections to the magnetic moments arise from the term
\eq
\label\Ncorr
\del\scr L = \frac {a_1 e}{N\Lam} v_\mu \twi F^{\mu\nu}
\rbra{\scr B} \op{\scr Q} \op{\sig_\nu} \rket{\scr B}.
\eeq
Because matrix elements of the operator $\op{\scr Q} \op{\sig^\mu}$ can be
$O(N)$, the contributions to the magnetic moments arising from this term
are $O(1)$ in the large-$N$ limit.
Including this term, the Coleman--Glashow relations in eq.\ \lorels\ still
hold, but the quark-model relations eqs.\ \QM\ and \QMtoo\ no longer hold.
Note that the relation eq.\ \QM\ involves states with zero
strangeness, and therefore receives further corrections only from isospin
breaking, which are expected to be about $5\%$.
The experimental deviation of this relation is therefore a direct measure of
the $1/N$ corrections, and we have no understanding of why the these
corrections are so small in this case.

\vfill\eject
\subsection{$SU(N_F)$-breaking Corrections}

In chiral perturbation theory, the leading $SU(N_F)$-breaking corrections
generally arise from loop graphs with PNGB intermediate states.
Such graphs can give nonanalytic dependence on the quark masses.
In the case of the magnetic moments, the leading dependence on $m_s$ is
$\sim m_s^{1/2}$ and $\sim m_s \ln m_s$
\ref\Pagels{D. G. Caldi and H. Pagels, \PRD{10}{3739}{1974}.}.
The diagrams which give rise to these corrections are shown in fig.\ 1.

The graph which gives rise to the $\sim m_s^{1/2}$ corrections is easily
evaluated using the meson--baryon coupling from eq.\ \lzero:
\eq
\label\graph
\eqalign{
{\rm fig.\ 1a} = -\frac{g^2 e}{2 f^2}\, &
\rbra{B'} \op{T_A \sig^\nu} \op{T_B \sig^\lam} \rket{B}
\,\tr(T_A [T_B, \scr Q]) \cr
& \times \myint \frac{d^4 k}{(2\pi)^4}\,
\frac{(2k + q)_\mu k_\nu (k + q)_\lam}
{(k^2 - M_A^2 + i0^{+})((k + q)^2 - M_B^2 + i0^{+})(k\cdot v + i0^{+})}, \cr}
\eeq
where the $T_A$ are $SU(N_F)$ generators normalized so that
$\tr(T_A T_B) = \delta_{AB}$, and
\eq
M^2_{AB} = \frac{\beta f}2 \tr(\{T_A, T_B\} m_q)
\eeq
is the mass-squared matrix of the PNGB's (see eq.\ \mesonL).
We have neglected the $O(m_s)$ mass differences between baryons in the same
$SU(N_F)$ multiplet, as well as the $O(1/N)$ mass differences between octet
and decuplet baryons.
(Including these effects gives corrections suppressed by $m_s$ and/or
$1/N$.)
The sum over all intermediate spin states is then included in eq.\ \graph,
with the large-$N$ relations properly taken into account.
Evaluating eq.\ \graph\ gives rise to a contribution to the magnetic moments
\eq
\del\mu^{\vphantom j}_{B' B} \sig^j_{B' B} =
\frac{g^2 e M_K}{16\pi f^2} \rbra{B'} \scr O_K^j \rket{B},
\eeq
where
\eq
\label\Kop
\scr O^j_K = (N + N_F)\, \tr[Q(1 - S)]\, \op{S \sig^j}
- \op{S} \op{\scr Q \sig^j} + \op{\scr Q} \op{S \sig^j}.
\eeq
This gives contributions to the magnetic moments which are
$O(N m_s^{1/2}) = O(N \ep^{1/2})$.

Including these contributions along with the leading term in eq.\ \leading,
we obtain 8 relations valid to $O(N \ep^{1/2})$ which are independent of
$g$.
One of these is the quark-model relation eq.\ \QM.
The remaining relations can be written
\eq
\eqalign{
\mu_{\Sig^-} + \mu_n + \mu_p &= 0, \quad (10\%) \cr
\mu_{\Xi^0} - 2\mu_{\Xi^-} &= 0, \quad (4\%) \cr
\mu_{\Sig^+} - 2\sqrt{3} \mu_{\Sig^0\Lam} + \mu_p &= 0, \quad (7\%) \cr
\mu_{\Sig^+} + \frac 1{\sqrt{3}} \mu_{\Sig^0\Lam} + \mu_\Lam + \mu_p &= 0,
\quad (2\%) \cr
\mu_{\Xi^-} + \mu_{\Sig^+} - \mu_{\Sig^-} - \mu_p &= 0, \quad (4\%) \cr
\mu_{\Om^-} - \mu_{\Xi^0} - \mu_{\Xi^-} &= 0. \quad (2\%) \cr}
\eeq
In addition, there is one relation which depends on $g$,
\eq
\label\gdep
\mu_{\Sig^-} - \mu_{\Xi^-} = \frac{g^2 e M_K}{8\pi f^2}\qquad
(-0.51 = -2.0).
\eeq
(We use $f = f_K \simeq 114 \MeV$ in the evaluation.)
The relations which are independent of $g$ work much better than the
$g$-dependent relations:
a fit including the $O(N m_s^{1/2})$ corrections and treating $g$ as a free
parameter has an average deviation of $0.08\ \mn$.
\ignore
(If we evaluate the operator in eq.\ \Kop\ assuming $N_F \ll N$, the average
deviation for this case is $0.09\ \mn$.)
\endignore
The nonanalytic corrections have the right sign, but their predicted
magnitude for the lowest-order value $g \simeq 0.8$ is too large.
However, we expect that including the $SU(N_F)$-breaking corrections
in the fit to the semi-leptonic decays will substantially {\it decrease} $g$
\heavyB\bigcorr, and it is not clear to us that the large discrepancy in
eq.\ \gdep\ indicates a breakdown of the expansion.

The vertex graph (fig.\ 1b) and wavefunction graphs (fig.\ 1c) combine to
give the contribution
\eq
\label\log
\eqalign{
{\rm figs.\ 1b,c} = \frac{i e g^2}{4 f^2} &
\ep_{\mu\nu\lam\rho} v^\nu q^\lam\,
\rbra{B'} \bigl[\op{T_A \sig^\al}, \left[\op{T_A \sig^\beta},
    \op{\scr Q \sig^\rho}\right]\bigr] \rket B \cr
& \times \myint \frac{d^4 k}{(2\pi)^4}\,
\frac{k_\al k_\beta}
{(k^2 - M_A^2 + i0^{+})(k\cdot v + i0^{+})((k - q)\cdot v + i0^{+})}. \cr}
\eeq
The double commutator can be written as a 1-body operator;
this contribution is therefore $\sim 1/N$ times a 1-body operator, and can
be at most $O(m_s \ln m_s) = O(N \ep^2 \ln \ep)$.
This is negligible compared to the $O(N \ep)$ counterterms which we will
discuss below.
The graph in fig.\ 1d gives a contribution which is $1/N$ times a 1-body
operator, which is negligible for the same reason as for eq.\ \log;
the graph in fig.\ 1e gives no contribution to the magnetic moments

The leading $SU(N_F)$-violating counterterms are
\eq
\label\NFcorr
\del\scr L = e v_\mu \twi F^{\mu\nu}
\rbra{\scr B} \biggl[
\frac{b_1}{\Lam^2} \op{(\scr Q m) \sig_\nu}
+ \frac{b_2}{N\Lam^2} \op{\scr Q \sig_\nu} \op{m}
+ \frac{b_3}{N\Lam^2} \op{\scr Q} \op{m \sig_\nu}
\biggl] \rket{\scr B}.
\eeq
These counterterms give $O(N m_s) = O(N \ep)$ contributions to the magnetic
moments.
\ignore
There are also $m_s$-dependent terms such as
\eq
\label\sigterm
\del\scr L = \frac{e b_4}{N \Lam^2} \tr(m)
v^\mu \ep_{\mu\nu\lam\rho} F^{\nu\lam}
\rbra{\scr B} \rbra{\scr B} \op{\scr Q \sig^\rho} \rket{\scr B},
\eeq
which do not give rise to $SU(N_F)$ violation in the magnetic moments.
(The reason for the extra factor of $1/N$ in this term is explained in
ref.\ \us.)
For the purposes of this paper, these terms can be absorbed into the
$SU(N_F)$-symmetric terms in eqs.\ \leading\ and \Ncorr.
\endignore
There are also $O(N m_s)$ contributions from the loop graph in fig.\ 1a, but
these have the same spin--flavor dependence as the counterterms considered
above, so we will not need to evaluate them explicitly.
Including the counterterms in eq.\ \NFcorr\ and the nonanalytic corrections
computed above, we obtain the relations
\eq
\eqalign{
\mu_{\Xi^0} + 2 \mu_{\Sig^+} + 2 \mu_{\Sig^-} + \mu_n &= 0, \quad (10\%) \cr
\mu_{\Xi^-} + 4 \mu_{\Sig^-} + 2\sqrt{3} \mu_{\Sig^0 \Lam} + 5 \mu_p
+ 8 \mu_n &= 0, \quad (2\%) \cr
\mu_{\Om^-} + 4 \mu_{\Xi^0} - 3 \mu_{\Xi^-} + 8 \mu_{\Sig^+}
+ 5 \mu_{\Sig^-} - 3 \mu_p + \mu_n &= 0, \quad (8\%) \cr
4 \mu_{\Xi^0} - \mu_{\Sig^+} - \mu_{\Sig^-} + 4\sqrt{3} \mu_{\Sig^0 \Lam}
- 6 \mu_\Lam + 4 \mu_n &= 0, \quad (6\%). \cr}
\eeq
The last of these relations was noted in
ref.\ \ref\cptmag{E. Jenkins, M. Luke, A. V. Manohar, and M. J. Savage,
\PLB{302}{482}{1993}.},
and is valid to $O(m_s)$ independent of the $1/N$ expansion.
The average deviation of a fit which treats all of the counterterm couplings
as free parameters is $0.08\ \mn$, the same as the fit including only the
$O(N m_s^{1/2})$ terms.

\section{Predictions}

We can use the results obtained above to predict the unmeasured decuplet
magnetic moments including all contributions up to $O(N \ep)$.
These may be measured in the future at CEBAF.

Up until now, the fits were used only to give a rough idea of how well the
expansion works.
We now use the fit to the magnetic moment data to make predictions, so
we give some details on the fit and the treatment of errors.
We add a theoretical uncertainty of $0.1\ \mu_N$ in quadrature with the
experimental error to obtain the error on the individual magnetic moments,
and then perform a $\chi^2$ fit.
This theoretical uncertainty is approximately the average deviation of the
fit when corrections up to $O(N\ep)$ are included, and is consistent with
the expectations of dimensional analysis:
the largest contributions not included are $m_s^{3/2}$ nonanalytic terms
\eq
\del\mu \sim \frac{M_K^3 M^{\vphantom 3}_N}{16\pi f^2 \Lam^2}
\mu^{\vphantom j}_N \sim 0.2\ \mn.
\eeq
We then obtain the predictions (in Bohr magnetons)
\eq
\label\decpred
\eqalign{
\mu_{\Del^{++}} &= 5.9 \pm 0.4, \cr
\mu_{\Del^+} = -\mu_{\Del^-} &= 2.9 \pm 0.2, \cr
\mu_{\Sig^{*+}} &= 3.3 \pm 0.2, \cr
\mu_{\Sig^{*0}} &= 0.3 \pm 0.1, \cr
\mu_{\Sig^{*-}} &= -2.8 \pm 0.3, \cr
\mu_{\Xi^{*0}} &= 0.65 \pm 0.2, \cr
\mu_{\Xi^{*-}} &= -2.3 \pm 0.15, \cr}
\eeq
The error quoted is purely the formal error from the fit.
Note that in the limit of exact $SU(3)$ flavor symmetry, the decuplet
magnetic moments are proportional to their charges.
Eq.\ \decpred\ gives definite predictions for the pattern of $SU(3)$
violation which would be very interesting to test.

We make no prediction for the $\Del^0$ magnetic moment.
The leading contribution to this magnetic moment comes from terms such as
\eq
\del\scr L = \frac{e b_4}{N \Lam^2} \tr(m \scr Q)\,
v^\mu \ep_{\mu\nu\lam\rho} F^{\nu\lam}
\rbra{\scr B} \op{\sig^\rho} \rket{\scr B},
\eeq
where $b_4 \sim 1$.
(The reason for the factor of $1/N$ in the coefficient is that the quark
diagrams which contribute to this term contain a fermion loop;
see refs.\ \us\largeNF.)
This gives a contribution to the magnetic moments $O(1/N) = O(N\ep^2)$,
which is higher order than the terms we are keeping.

Decuplet magnetic moments were also considered in a chiral expansion in
ref.\ \ref\BSS{M. N. Butler, M. J. Savage, and R. P. Springer,
UCSD/PTH 93-22, QUSTH-93-05, Duke-TH-93-56, hep-ph/9308317.}.
This paper computes $O(m_s^{1/2})$ and $O(m_s \ln m_s)$
$SU(3)$-violating contributions to the decuplet magnetic moments without
making use of the $1/N$ expansion.
Their predictions differ considerably from ours:
for example, their predicted values for $\mu_{\Sig^{*0}}$ and
$\mu_{\Xi^{*0}}$ are significantly smaller than ours.
It is worth noting that they do not include $O(m_s)$ counterterm
contributions, which are not expected to be significantly smaller than the
$O(m_s \ln m_s)$ which they compute, whereas the largest terms which are
omitted in our analysis are suppressed by a {\it power} of the expansion
parameter.
We hope that experiment will be able to decide between these different
predictions in the future.

The radiative decays of decuplet baryons also receive contributions from
the same terms that give rise to the magnetic moments.
However, the momentum transfer for the process $T \to B\gam$ is
\eq
|\vec q| \sim M_T - M_B \sim \frac{\Lam}{N},
\eeq
and the loop graph in fig.~1a has nontrivial dependence on
$|\vec q| / (M_T - M_B) \sim 1$.
This makes the coefficients of the effective operators in the expansion
different from the ones appearing in the expansion of the magnetic moments.
In principle, this difference is computable, but it depends on the value
of the axial coupling $g$, which we know is not well determined.
We will return to these issues in a future publication.

\ignore
The terms discussed above give rise to magnetic moment ($M1$) transitions.
There are also electric quadrupole ($E2$) transitions mediated by terms
such as
\eq
\del\scr L_{E2} = \frac{c}{N\Lam^2} \partial^\mu F^{\nu\lam} v_\lam
\rbra{\scr B} \op{\scr Q \sig_\mu} \op{\sig_\nu} \rket{scr B}.
\eeq
The ratio of $E2$ to $M1$ amplitudes is therefore
\eq
\delta_{E2/M1} \sim \frac 1N\, \frac{M_T - M_B}{\Lam} \sim \frac 1{N^2},
\eeq
and we can neglect the $E2$ amplitude.
Experimentally, it is indeed found that $\delta_{E2/M1} \ll 1$
\ref\PDG{Particle Data Group, \PRD{45}{Part II}{1992}.}.
The radiative decay width due to the leading $M1$ term in eq.\ \leading\ is
given by
\eq
\Gam(T \to B\gam) = \frac{a_0^2 e^2}{\pi \Lam^2}\,
\frac{M_T}{M_B} \left( \frac{M_T^2 - M_B^2}{2M_T} \right)^3
\scr C_{TB},
\eeq
where $\scr C_{TB}$ is defined by
\eq
\sum_{\rm spins} \rbra T \op{\scr Q \sig^j} \rket B
\rbra B \op{\scr Q \sig^k} \rket T = \scr C_{TB} \del^{jk}.
\eeq
For the measured decay $\Del^+ \to p\gam$, the $SU(3)$-violating terms in
eqs.\ \Ncorr\ and \NFcorr\ do not contribute, and the predicted width is
\eq
\Gam(\Del^+ \to p\gam) = 0.39 \pm 0.03 \MeV.
\eeq
This disagrees badly with the measured width $0.74$--$0.86 \MeV$ \PDG.
\endignore

\section{Acknowledgements}

We would like to thank M. J. Savage and R. Sundrum for discussions.
J. M.--R. acknowledges the support of a DOE distinguished postdoctoral
research fellowship, and M. W. acknowledges the support of an SSC fellowship
from the TNRLC.
This work was supported by the Director, Office of Energy Research, Office
of High Energy and Nuclear Physics, Division of High  Energy Physics of the
U.S. Department of Energy under Contract DE-AC03-76SF00098.

\listrefs

\vfill\eject
\centerline{\bf Figure Captions}
\vskip .2in
\noindent Fig.\ 1.
Feynman graphs giving rise to nonanalytic corrections to the baryon magnetic
moments.
Fig.~1c is wavefunction renormalization;
fig.~1e does not contribute to the magnetic moments.

\bye